\documentclass[nofootinbib, reprint, preprintnumbers, amsmath, amssymb, amsfonts, aps, pra, superscriptaddress, floatfix]{revtex4-2}
\usepackage[utf8]{inputenc}
\usepackage{mathtools}
\usepackage{graphicx}
\usepackage{dcolumn}
\usepackage{bm}
\usepackage{physics}
\usepackage{color}
\usepackage{braket}
\usepackage[normalem]{ulem}
\usepackage{enumerate}
\usepackage{enumitem}
\usepackage[colorlinks=true,linkcolor=blue,citecolor=blue,urlcolor =blue]{hyperref}
\usepackage{mathrsfs}
\usepackage{bbm}

\usepackage{comment}
\usepackage{lipsum}

\begin{document}
\title{Expressive Quantum Supervised Machine Learning using Kerr-nonlinear Parametric Oscillators}

\author{Yuichiro Mori}
\email{mori-yuichiro.9302@aist.go.jp}
\affiliation{Global Research and Development Center for Business by Quantum-AI Technology (G-QuAT), National Institute of Advanced Industrial Science and Technology (AIST), 1-1-1, Umezono, Tsukuba, Ibaraki 305-8568, Japan}%

\author{Kouhei Nakaji}
\email{kohei.nakaji@utoronto.ca}
\affiliation{
    Chemical Physics Theory Group, Department of Chemistry, University of Toronto, Toronto, Ontario, Canada
}
\affiliation{Global Research and Development Center for Business by Quantum-AI Technology (G-QuAT), National Institute of Advanced Industrial Science and Technology (AIST), 1-1-1, Umezono, Tsukuba, Ibaraki 305-8568, Japan}
\affiliation{%
    Quantum Computing Center, Keio University, 3-14-1 Hiyoshi, Kohoku-ku, Yokohama, Kanagawa, 223-8522, Japan}

\author{Yuichiro Matsuzaki}
%\email{matsuzaki.yuichiro@aist.go.jp}
\email[Present E-mail address: ]{ymatsuzaki872@g.chuo-u.ac.jp}
\affiliation{Global Research and Development Center for Business by Quantum-AI Technology (G-QuAT), National Institute of Advanced Industrial Science and Technology (AIST), 1-1-1, Umezono, Tsukuba, Ibaraki 305-8568, Japan}%
\affiliation{NEC-AIST Quantum Technology Cooperative Research Laboratory,
National Institute of Advanced Industrial Science and Technology (AIST), Tsukuba, Ibaraki 305-8568, Japan}

\author{Shiro Kawabata}
\email{s-kawabata@aist.go.jp}
\affiliation{Global Research and Development Center for Business by Quantum-AI Technology (G-QuAT), National Institute of Advanced Industrial Science and Technology (AIST), 1-1-1, Umezono, Tsukuba, Ibaraki 305-8568, Japan}%
\affiliation{NEC-AIST Quantum Technology Cooperative Research Laboratory,
National Institute of Advanced Industrial Science and Technology (AIST), Tsukuba, Ibaraki 305-8568, Japan}

\date{\today}

%%%%%%%%
\begin{abstract}
Quantum machine learning with variational quantum algorithms (VQA) has been actively investigated as a practical algorithm in the noisy intermediate-scale quantum (NISQ) era. 
Recent researches reveal that the data reuploading, which repeatedly encode classical data into quantum circuit, is necessary for obtaining the expressive quantum machine learning model in the conventional quantum computing architecture. However, the data reuploding tends to require large amount of quantum resources, which motivates us to find an alternative strategy for realizing the expressive quantum machine learning efficiently. 
In this paper, we propose quantum machine learning with
Kerr-nonlinear Parametric Oscillators (KPOs), as another promising quantum computing device. We use not only the ground state and first excited state but also use higher excited states, which allows us to use a large Hilbert space even if we have a single KPO. 
Our numerical simulations show that the expressibility of our method with only one mode of the KPO is much higher than that of the conventional method with six qubits. 
Our results pave the way towards resource efficient quantum machine learning, which is essential for the practical applications in the NISQ era.
\end{abstract}
%%%%%%%%%

\maketitle

%%%%%%%%%
\section{Introduction}
\label{se:intro}
%%%%%%%%%
The quantum computers have attracted much attention due to its potential impact on quantum chemistry~\cite{Aspuru2005_science, Cao2019_Chemrev, McArdle2020_revmodp, Armaos2020APA}, machine learning~\cite{Schuld2015_Contphys, Biamonte2017_nature, Schuld2022PRXQ}, cryptography~\cite{Shor1994_Proc, Shor_1997SIAM, Lenstra2000_Des}, search problems~\cite{Grover1996_proc} and so on. With advancements in quantum technology, commercially available quantum computers have become a reality. In principle, we could realize a fault-tolerant quantum computer, if the number of qubits is more than $10$ millions with a fidelity around 0.999~\cite{Cody_2012PRX, DSimon_2013natcom, Gidney_2021}.
However, in the current device, the available number of qubits is an order of $500$ or less, which is much smaller than that required for the fault-tolerant quantum computation.
A more feasible scenario to be realized in the near future is the so-called NISQ regime~\cite{Preskill_2018quant, Bharti_2022RMP}.

Numerous quantum algorithms have been designed for execution on NISQ devices. Among these, VQAs are considered some of the most promising applications for NISQ devices~\cite{Bharti_2022RMP, Endo2021_JPSJrv}. Specifically, quantum machine learning has emerged as an appealing use case for VQAs. As a NISQ algorithm, quantum machine learning has been predominantly investigated in the context of qubit-based systems.
Recent studies have shown that data reuploading, the process of repeatedly encoding classical data into quantum circuits, is essential for achieving expressive quantum machine learning models within traditional quantum computing frameworks~\cite{perez2020data,gil2020input,Schuld_2021_PRA}. However, data reuploading often demands much quantum resources.
This encourages us to seek alternative approaches to achieve expressive quantum machine learning. We could adopt a photonic device where fock states can be used and the necessary quantum resources for the data embedding with this device could be different from that with the conventional approach using qubits~\cite{Killoran_2019PRR, Steinbrecher2019_npjq, Volkoff2021_RusLas, Gan2022_EPJQ, Liu2023_MLST}.

On the other hand, the KPO is one of the candidates to realize quantum computation~\cite{milburn1991quantum,wielinga1993quantum,Cochrane1999_PRA}. 
The KPO is a parametric oscillator with large Kerr nonlinearity. This Kerr nonlinearity can be used to generate cat states. We can realize the KPO by using superconducting resonators with Josephson junctions~\cite{Bourassa2012PRA, Meaney2014EPJQ}.
The KPO is one of the candidates to perform gate-type quantum computation~\cite{Cochrane1999_PRA,Goto_2016PRAR,Puri2017_npjq}
and quantum annealing~\cite{Goto_2016srep, Puri2017_NatCom}, and the KPO qubit is realized experimentally~\cite{Grimm2020_nature}. It is known that the KPO qubit is highly tolerant to bit-flip errors and we can exploit this property to reduce the overhead for fault-tolerant quantum computation~\cite{Puri2017_npjq, Masuda_2022PRAppl}.

In this paper, we propose to use the KPO for the supervised machine learning with variational algorithm. KPO is a bosonic system, and we can in principle use the infinitely large Hilbert space with the single KPO.
Also, unlike the conventional approach to use parametrized gates, we use a natural Hamiltonian dynamics where we change the Hamiltonian parameter to implement the variational algorithm. We numerically study the performance of our method to use the KPO with that of the conventional method with qubits.

In our method, we start from a coherent state with an amplitude of $\alpha$.
Importantly, we numerically find that, by changing the amplitude, we can tune the expressibility. 
Since we encode the input classical data by using the detuning of the KPO, we can include higher frequency as we increase the amplitude of the coherent state. We expect that the high frequency terms will improve the expressibility, and we confirm this point by using numerical simulations.
As the expressivity increases, on the other hand, more often overfitting occurs, and so our method allows us to optimize the expressibility by tuning the amplitude of the coherent state.

This paper is organized as follows. In Sec.~\ref{sec:KPO}, we review the physics of single and multiple KPO systems. The latter is called KPO network. In Sec.~\ref{sup_mac_lea}, we explain a standard supervised machine learning algorithm as a NISQ algorithm, and a supervised machine learning algorithm for KPO is proposed based on the ideas in Sec.~\ref{sec:sqmlwKPO}. We performed numerical simulations to validate our proposing method. In Sec.~\ref{sec:results}, we explain the simulations, and the results precisely. Finally, we conclude with some final thoughts in Sec.~\ref{sec:conclusion_and_discussion}.

%%%%%%%%%%%%%%%%%%
\section{KPO}
\label{sec:KPO}
%%%%%%%%%%%%%%%%%%%%
KPO is a bosonic system with a nonlinear effect called Kerr nonlinearity. 
Here we first describe a single KPO, and next explain a network of KPOs that have been used for a gate-type quantum computer or quantum annealing.

First, in a frame rotating at half the pump frequency of the parametric drive and in the rotating-wave approximation, the Hamiltonian of the single KPO is written as~\cite{Goto_2016srep, Goto_2019JPSJ}
\begin{align}
    \hat{H}&=\chi \hat{a}^{\dagger 2}\hat{a}^{2} + \Delta \hat{a}^{\dagger} \hat{a}\nonumber \\
    &\quad- p(\hat{a}^{2} + \hat{a}^{\dagger 2}) + r(\hat{a}+\hat{a}^{\dagger}),\label{def:KPOHamil}
\end{align}
where $\chi$, $\Delta$, $p$ and $r$ are the Kerr nonlinearity, the detuning, the pump amplitude of the parametric drive, and the strength of the coherent drive, respectively. 

We can easily tune $\Delta$, $p$, and $r$ during the experiment by changing the parameters of the external driving fields. 
Although we can tune $\chi$ by changing magnetic flux penetrating the superconducting loop of the KPO, the dynamic range is typically small, and therefore we assume that 
$\chi$ is fixed at a specific value.

The coherent state is defined by
\begin{align}
\ket{\alpha} = e^{-\frac{|\alpha|^{2}}{2}}\sum_{k} \frac{\alpha^{k}}{\sqrt{k!}}\ket{k}, \label{eq:coherent_state}
\end{align}
where $\ket{k}$ are the fock states. 
The system is initially prepared in the coherent state in our method. 
For a linear resonator, we can prepare the coherent state by adding the coherent driving term $r(\hat{a}+\hat{a}^{\dagger})$.
However, due to the term $\chi \hat{a}^{\dagger 2}\hat{a}^{2}$ in Eq.~\eqref{def:KPOHamil}, we cannot prepare the coherent state just by adding the coherent drive. 
Instead, we can prepare the coherent state by using the KPO as follows. 
By setting $p=r=0$, the Hamiltonian \eqref{def:KPOHamil} becomes
\begin{align}
\hat{H}&=\chi \hat{a}^{\dagger 2}\hat{a}^{2} + \Delta \hat{a}^{\dagger} \hat{a}.\label{def:KPOHamil_przero}
\end{align}
If $\Delta >\chi$ is satisfied, the ground state of this Hamiltonian becomes the vacuum state $\ket{0}$. On the other hand, when $\Delta$ and $r$ are zero, Eq.~\eqref{def:KPOHamil} can be rewritten as
\begin{align}
\hat{H} =\chi\left(\hat{a}^{\dagger 2}-\frac{p}{\chi}\right)\left(\hat{a}^{2}-\frac{p}{\chi}\right)-\frac{p^{2}}{\chi}, \label{eq:nodetuning}
\end{align}
and the ground state is in the eigenspace that is spanned by two coherent states $\ket{\sqrt{p/\chi}}$ and $\ket{-\sqrt{p/\chi}}$. 
By adding a coherent drive as a perturbation, we can solve the degeneracy, and the ground state becomes approximately $\ket{\sqrt{p/\chi}}$ with a negative value of $r$. 
If we prepare a vacuum state with the Hamiltonian of Eq.~\eqref{def:KPOHamil_przero}, 
the system is in the ground state. By adiabatically changing the Hamiltonian from Eq.~\eqref{def:KPOHamil_przero} to Eq.~\eqref{eq:nodetuning}, we obtain the coherent state $\ket{\sqrt{p/\chi}}$ due to the adiabatic theorem. This operation is frequently used in quantum annealing with KPOs~\cite{Goto_2016PRAR,Goto_2016srep,Goto_2019JPSJ,Puri2017_npjq}.

Next, the Hamiltonian of multiple KPOs, which is called a KPO network is written as
\begin{align}
\hat{H} = &\sum_{j=1}^{K} \chi_{j}\hat{a}^{\dagger 2}_{j}\hat{a}_{j}^{2} + \Delta_{j}\hat{a}^{\dagger}_{j}\hat{a}_{j}\nonumber \\
&\qquad\quad -p_{j}(\hat{a}_{j}^{2}+\hat{a}^{\dagger 2}_{j})+r_{j}(\hat{a}_{j} + \hat{a}^{\dagger}_{j})\nonumber\\
&\ +\sum_{j>j'}^{K} \left(J_{jj'}\hat{a}^{\dagger}_{j}\hat{a}_{j'}+J_{jj'}^{\ast}\hat{a}^{\dagger}_{j'}\hat{a}_{j}\right). \label{eq:KPO_network_Hamil}
\end{align}
where $K$ denotes the number of KPOs and $J_{jj'}$ denotes the coupling strength between KPOs. 
Here, we assume that we fix the values of $\chi_{j}$ and $J_{jj'}$ during the experiment, while we can control the values of $\Delta_{j}$, $p_{j}$, and $r_{j}$. 

If $J_{jj'}$ is zero, we can independently perform the adiabatic state preparation described above, and prepare the following state.
\begin{align}
\bigotimes_{j=1}^{K} \ket{\alpha_{j}}\label{eq:separable_state}
\end{align}
Here, each $\alpha_{j}$ is the eigenvalue of $\ket{\alpha_{j}}$ with the annihilation operator on the $j$-th KPO $\hat{a}_{j}$. 

It is worth mentioning that even when $J_{jj'}$ is nonzero, we can prepare the product of the coherent state as follows.
Let us assume that $\Delta _j$, $r_j$, and $J_{ij}$ are much smaller than $p_j$ and $\chi_j$. In this case, the last terms of the Hamiltonian Eq.~\eqref{eq:KPO_network_Hamil} can be interpreted as the longitudinal-field Ising Hamiltonian in a coherent-state basis.
If $J_{ij}$ is negative, we have a ferromagnetic Hamiltonian.
Moreover, by setting $J_{ij}$ to be much smaller than $r_j$, the state in Eq.~\eqref{eq:separable_state} can be a ground state, and so we can prepare this state in an adiabatic way. Also, a coupling scheme of KPOs with high fidelity has already been proposed theoretically~\cite{Goto_2019JPSJ, Masuda_2022PRAppl, aoki2023control}.

%%%%%%%%%%%%%%%%%%%%%
\section{Quantum supervised machine learning as a NISQ algorithm}
\label{sup_mac_lea}
%%%%%%%%%%%%%%%%%%%%%
In this section, let us review a quantum supervised machine learning as a preparation for introducing our model. In a supervised learning task, a number of training set $\{(\boldsymbol{x}_{m}, \boldsymbol{y}_{m})\}_{m=1}^N$ is given. Here, all input data $\boldsymbol{x}_{m}$ (output data $\boldsymbol{y}_{m}$) are $d_{x}$ ($d_{y}$) dimensional arrays.
Suppose that there is a hidden relationship between an input data $\boldsymbol{x}$ and the output data $\boldsymbol{y}$ as $\boldsymbol{y} = \tilde{f}(\boldsymbol{x})$ with a function $\tilde{f}$. The objective of the task is to find the hidden relationship $\tilde{f}$ from the training data. More specifically, we define the model function $f$ and optimize it so that it becomes close to $\tilde{f}$ by using the training data.

In most of the quantum machine learning with near-term devices, we use a parameterized quantum circuit to construct a model function. More precisely, by tuning a parameter, we try to minimize a cost function. In usual cases, we choose the mean squared error
\begin{align}
L(\boldsymbol{\theta}) = \frac{1}{N}\sum_{m=1}^{N} \left|f(\boldsymbol{x}_{m};\boldsymbol{\theta})-\boldsymbol{y}_{m}\right|^{2},
\end{align}
for the cost function. Here, $N$ is the number of data sets, $f(\boldsymbol{x};\boldsymbol{\theta})$ is an array as an output of the parameterized quantum circuit, and $\boldsymbol{\theta}$ is the corresponding parameter.
Let us summarize such a quantum machine learning as follows.
\begin{enumerate}
\item Prepare an initial state $\ket{\psi}$, and apply an input gate $\hat{U}(\boldsymbol{x})$ to encode the input data $\{\boldsymbol{x}_i\}$.
\item Apply a parameterized unitary $\hat{V}(\boldsymbol{\theta})$ to the state.
\item Measure the expectation values of an observable $\hat{M}$, and we define the function as $f(\boldsymbol{x};\boldsymbol{\theta})= \langle \hat{M}\rangle $.
\item  By repeating the above three steps, minimize the cost function $L$ by tuning the parameter $\boldsymbol{\theta}$ iteratively.
\end{enumerate}

The function $f(\boldsymbol{x})$ is represented as
\begin{align}
f(\boldsymbol{x};\boldsymbol{\theta})=\bra{\psi}\hat{U}^{\dagger}(\boldsymbol{x})\hat{V}^{\dagger}(\boldsymbol{\theta})\hat{M}\hat{V}(\boldsymbol{\theta})\hat{U}(\boldsymbol{x})\ket{\psi}.\label{eq:f_func}
\end{align}

According to a previous study~\cite{Schuld_2021_PRA}, we may not expect high expressibility with a parametrized quantum circuit using single-qubit rotations in the NISQ era. In fact, the study shows that only a sinusoidal curve can be obtained as Eq.~\eqref{eq:f_func} with using a qubit and single-qubit rotations and if we want different functions as an output, we need to prepare more qubits or obtain other outputs than Eq.~\eqref{eq:f_func} with adding another operation called data-reuploading. However, neither increasing the number of qubits nor increasing the number of gate operations that cause noise is desirable for the NISQ algorithm.

%%%%%%%%%%%%%%%%%%
\section{Quantum supervised machine learning with KPO}
\label{sec:sqmlwKPO}
%%%%%%%%%%%%%%%%%%
We introduce our method to use the KPO for supervised quantum machine learning. We begin by describing a simplified scenario with $d_x=d_y=1$ by using the single KPO. Next, we explain how to use the KPO network for supervised quantum machine learning with $d_x=d_y=1$. Finally, we describe a scenario to implement supervised quantum machine learning with $d_x>1$ and/or $d_y>1$ by using the KPO network.

%%%%
\subsection{$d_{x} = d_{y} =1$ case}
\label{subsec:dxdy1}
%%%%

%%%%
\subsubsection{Single KPO}
\label{subsubsec:single-mode}
%%%%
In our paper, the initial state is set to be a coherent state. 
Also, to upload the classical data, we could adopt $\hat{U}(x) = e^{-i\pi x \hat{n}}$, where $\hat{n}$ is the number operator defined by $\hat{n}=\hat{a}^{\dagger}\hat{a}$. However, if we use the KPO, it is difficult to realize in situ tunability of the nonlinearity $\chi$. Assuming that we fix the value of $\chi$ during the experiment, we will adopt the following operator to upload the classical data 
\begin{align}
\hat{U}(x) = e^{-i\tilde{\chi}\hat{n}^{2}-i\pi x \hat{n}}, \label{Aembed_x}
\end{align}
where we have,
\begin{align}
\tilde{\chi}&=t_{d}\chi,\\
\pi x &= t_{d}(\Delta-\chi).
\end{align}
In the actual experiment, we can easily tune the time duration $t_{d}$ and detuning $\Delta$. Throughout of this paper, we fix the value of $\tilde{\chi}$.

Let us define a set of unitary operators $\hat{V}_{i}(\Delta_{i}, p_{i}, r_{i})$ 
\begin{align}
\hat{V}_{i}(\Delta_{i}, p_{i}, r_{i}) = e^{-i\tau \hat{H}}. \label{eq:unitary_vi}
\end{align}
where $\hat{H}$ denotes the Hamiltonian of the KPO~\eqref{def:KPOHamil} and $\tau$ denotes an evolution time by the Hamiltonian. By turning on and off the parameters of the Hamiltonian, we can construct a unitary operator 
\begin{align}
\hat{V}(\boldsymbol{\theta})=\prod_{i}^{D} \hat{V}_{i}(\Delta_{i}, p_{i}, r_{i}), \label{eq:prod_unitary}
\end{align}
where $D$ is the number of combinations of $(\Delta_{i}, p_{i}, r_{i})$. 
Here, $\boldsymbol{\theta}$ corresponds to a set of parameters $\{\Delta_{j}, p_{j}, r_{j}\}_{j=1}^{D}$. For simplicity, we define 
\begin{align}
\theta_{k} :=
\begin{cases}
\Delta_{i} & k = 3 i -2,\\
p_{i} & k = 3 i - 1,\\
r_{i} & k = 3 i,
\end{cases}
\end{align}
with $i = 1, \dots, d.$ 
We choose $\hat{a}+\hat{a}^{\dagger}=\hat{M}$ as the observable to be measured.
Since a bosonic system has an infinite dimensional Fock space, even a single KPO may have the ability to approximate the target function, while the previous approach required multiple qubits to represent the target function. 

To minimize the cost function, we need to tune the parameter $\boldsymbol{\theta}$. For this purpose, we should adopt a classical algorithm to show how we should update the parameters based on the expectation value of $\hat{M}$.

Several types of classical algorithms are used to update $\boldsymbol{\theta}$. One of them is the gradient descent method to use the gradient of the cost function. 
If we construct the unitary operator $\hat{V}(\boldsymbol{\theta})$ by using a sequence of parameterized gates, we can use the so-called parameter shift rule~\cite{Mitarai_2018_PRA, Wierichs_2021Quant} to determine the gradient. 
On the other hand,
since we use the Hamiltonian dynamics to realize the unitary operator $\hat{V}(\boldsymbol{\theta})$, 
it is not straight forward to use the parameter shift rule. 
We could use a numerical differentiation where we measure small changes in $f(x;\boldsymbol{\theta})$ when we incrementally increase $\boldsymbol{\theta}$.
by changing $\boldsymbol{\theta}$ in small increments and detecting the resulting small changes in output $f(x;\boldsymbol{\theta})$.
However, to detect the small changes, this method requires a large number of measurements. 

If we cannot use a sufficient number of shots, 
we could adopt an optimization using the Nelder-Mead or Powell method, which do not use the information of gradients. 
Throughout of this paper, we use the Nelder-Mead method~\cite{Nelder_1965CompJ} for our simulation.

Our method to use the single KPO needs to access higher excited states in the Fock space, which may cause experimental difficulties. This problem may be circumvented by using the KPO network.

\subsubsection{KPO network}
Next, we consider a case using a KPO network. We prepare the product state of the coherent state \eqref{eq:separable_state} as the initial state.
To upload the classical data, we apply the following operator on the $j$-th KPO,
\begin{align}
\hat{U}_{j}(x) = e^{-i\tilde{\chi}\hat{n}_{j}^{2}-i\pi x \hat{n}_{j}}, \label{Aembed_x_network}
\end{align}
where we have $\tilde{\chi}_j=t_{d}\chi_j$ and $\pi x= t_{d}\Delta_j.$ 
We define a unitary operator with $3K$ parameters, 
\begin{align}
\hat{V}(\vec{\Delta}, \vec{p}, \vec{r}) = e^{-i t_{d} \hat{H}}, \label{eq:multi-unitary}
\end{align}
where $\hat{H}$ is given by Eq.~\eqref{eq:KPO_network_Hamil}. 
Here, $\vec{\Delta}=(\Delta_1, \Delta_2, \cdots, \Delta _K)$, $\vec{p}=(p_1, p_2,\cdots ,p_{K})$ and $\vec{r}=(r_1,r_2,\cdots, r_K)$ are $K$ dimensional arrays. 

If we need more than $3K$ adjustable parameters, we 
could consider a different combination of $\vec{\Delta}$, $\vec{p}$, and $\vec{r}$. Let us define a set of such a combination as $\{\vec{\Delta}_{i}, \vec{p}_{i}, \vec{r}_{i}\}_{i=1}^{D}$ where $D$ is the number of the combination. 
Thus, we can generate $D$ different unitary operators based on Eq.~\eqref{eq:multi-unitary}. 
When we sequentially implement these, 
the unitary operator is given as
\begin{align}
\hat{V}(\boldsymbol{\theta}) = \prod_{i}^{D}\hat{V}_{i}(\vec{\Delta}_{i},\vec{p}_{i},\vec{r}_{i}).\label{eq:vtheta_multi}
\end{align} 
Here, $\boldsymbol{\theta}$ corresponds to a set of parameters as described below. 
\begin{align}
&(\vec{\Delta}_{1},\vec{p}_{1},\vec{r}_{1},...,\vec{\Delta}_{D},\vec{p}_{D},\vec{r}_{D})\nonumber\\
&\quad = (\Delta_{11},\Delta_{12},...,\Delta_{1K},p_{11},p_{12},...,p_{1K},...,r_{DK}). \nonumber
\end{align}
After applying $\hat{V}(\boldsymbol{\theta})$ given by Eq.~\eqref{eq:vtheta_multi}, we measure an observable $\hat{M}$. For this $\hat{M}$, we can choose an observable of $\hat{a}_{1}+\hat{a}^{\dagger}_{1}$, for example. 

%%%
\subsection{$d_x>1$ and/or $d_y>1$ case}
\label{subsec:dxdymoth2}
%%%
We describe our method to implement supervised quantum machine learning with 
$d_x>1$ and/or $d_y>1$ by using the KPO network.
Let us assume $K\geq d_x$. For $j=1,2,\cdots, d_x$,
we define
\begin{align}
\hat{U}_{j}(x_{j}) = e^{-i\tilde{\chi}\hat{n}_{j}^{2}-i\pi x_{j} \hat{n}_{j}}, \label{Aembed_xj}.
\end{align}
To upload the classical data, we use a unitary operator of
$\prod_{j=1}^{d_x}\hat{U}_{j}(x_{j})$.

Subsequently, we apply $\hat{V}(\boldsymbol{\theta})$ in Eq.~\eqref{eq:vtheta_multi}, and measure a set of observable $\{\hat{M}_{k}\}_{k=1}^{d_{y}}$. The expectation value of $\hat{M}_{k}$ corresponds to the $k$-th component of $\boldsymbol{y}$. By repeating these steps, we update the parameter $\boldsymbol{\theta}$ to minimize the cost function.
In principle, we could use the single KPO with $d_x>1$ and $d_y>1$, and we discuss such an example in Appendix~\ref{append:mulplefun1kpo}.

%%%%%%
\subsection{Potential advantage to use KPOs}
\label{subsec:merit}
%%%%%%
Even if we can use only a single KPO, the function obtained as Eq.~\eqref{eq:f_func} is expected to 
exhibit a large expressibility. Similar to the previous study~\cite{Schuld_2021_PRA}, we construct the Fourier spectrum of Eq.~\eqref{eq:f_func} in the case of a single KPO as $d_x=d_y=1$. 

When $\chi$ is negligibly small, 
we obtain
\begin{align}
    f(x;\boldsymbol{\theta}) &= \braket{\alpha|e^{i\pi x \hat{n}}\hat{V}^{\dagger}(\boldsymbol{\theta})\hat{M}\hat{V}(\boldsymbol{\theta})e^{-i\pi x \hat{n}}|\alpha} \nonumber \\
    &=e^{-|\alpha|^{2}}\sum_{k, l=0}^{\infty} \braket{k|\hat{V}^{\dagger}(\boldsymbol{\theta})\hat{M}\hat{V}(\boldsymbol{\theta})|l}\frac{\alpha^{l}\alpha^{\ast k}}{\sqrt{k! l!}}e^{i\pi x (k-l)},\label{eq:fourier_peak}
\end{align}
which is the Fourier series.
Importantly, there are high frequency terms in this form, and the number of terms is infinite. This would improve expressibility. A similar discussion has been made 
by Gan {\it{et al}} in the context of a multi-mode photonic device, which supports our claim~\cite{Gan2022_EPJQ}.

If we can provide an appropriate $\hat{V}(\boldsymbol{\theta})$ and $\hat{M}$, we could represent any function that can be represented by the Fourier series.
Moreover, previous research shows that the Kerr-nonlinearity could enhance the performance of a specific scheme of quantum machine learning~\cite{Liu2023_MLST}, and so our method to utilize the Kerr-nonlinearity might improve the expressibility.

On the other hand, if we use ordinary qubits, the number of high frequency terms is limited by the finite number of qubits. This could limit the expressibility, as suggested in~\cite{Schuld_2021_PRA}.
To improve the expressibility, we could increase the number of qubits~\cite{Schuld_2021_PRA} or circuit depth.
However, it is difficult to increase the number of qubits or circuit depth in the NISQ device.

%%%%%%%%%%%%%%
\section{Simulations and Results}
\label{sec:results}
%%%%%%%%%%%%%%
To evaluate the performance of our proposed method, we perform numerical simulations for $d_x=d_y=1$, and compare the results of our method with that of the conventional one~\cite{Mitarai_2018_PRA}. Specifically, we perform the fitting of $\tilde{f}(x)=$
$e^{-36x^{2}}$ (Gaussian), $|x|$, and $0.4\sin(4\pi x)+0.5\sin(6\pi x)$. 
Also, we perform the fitting of the square wave defined as
\begin{align}
    \tilde{f}(x) = \begin{cases}
        1 & (|x|<0.4) \\
        0 & (|x|\geq 0.4).
    \end{cases}
\end{align}

We create the training set as follows. We set $N=100$.
First, we randomly choose a value between $-1$ and $1$, and adopt these values as $x_m$. Next, for each $x_m$, we calculate $\tilde{f}(x_m)$ by using the given function $\tilde{f}$ and assign this value as $y_m$.

For our method to use a single KPO, we choose
$\chi=0.1$, $t_{d}=\tau=0.7$, $\hat{M} = \hat{a} + \hat{a}^{\dagger}$, and $D=12$. Also, we set the cut-off of the Hilbert-space dimension as $25$.

For the conventional method~\cite{Mitarai_2018_PRA}, we set depth $D=2$, the number of qubit $K=6$, time step $\tau=10$, and $\hat{M}=2Z^{(1)}$. Precise setups of the conventional method is given in Appendix~\ref{app:A}. Here, for a fair comparison, we set the number of parameter $\theta$ as $36$, which is equal to that of our method.

\begin{figure*}
    \includegraphics[width=11cm]{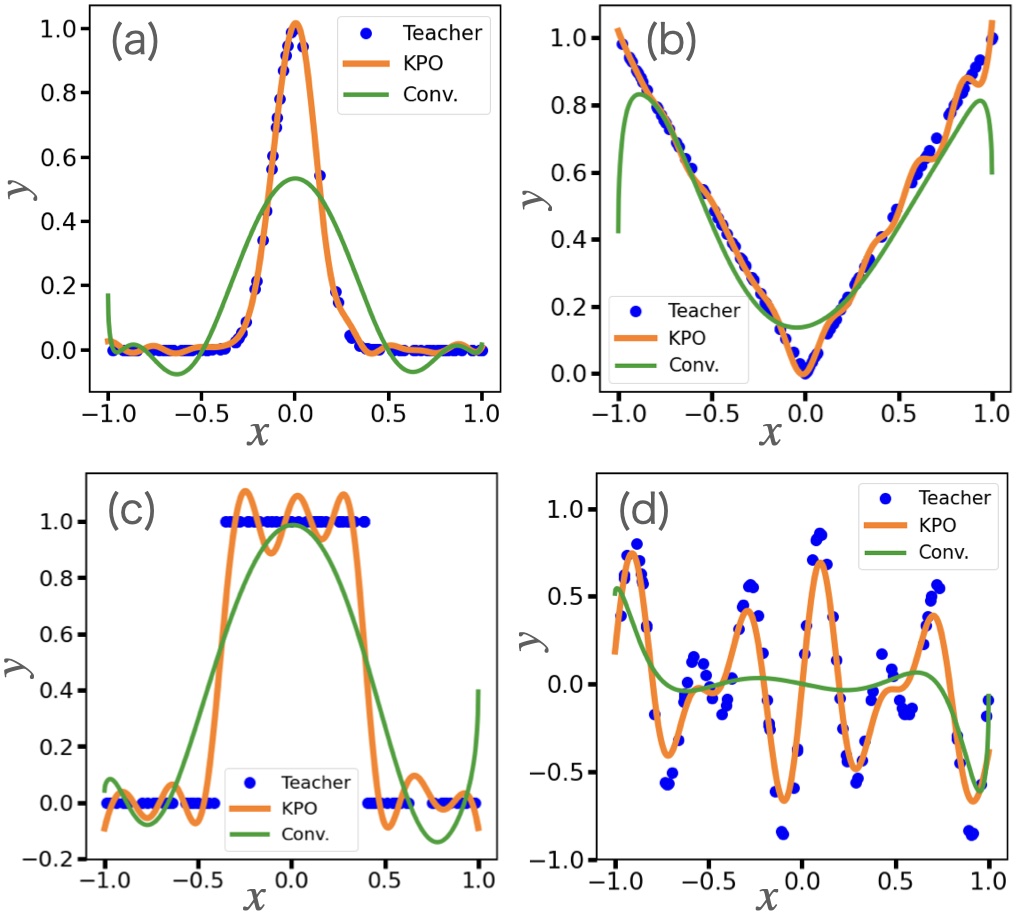}
    \caption{Demonstration of 
    quantum machine learning to represent functions.
    Blue dots indicate the teacher data. KPO (conventional) indicate the output by our (conventional) method after the optimization. We fit (a) $e^{-36x^{2}}$, (b) $|x|$, (c) Square wave and (d) $0.4\sin(4\pi x)+0.5\sin(6\pi x)$, respectively.}
    \label{fig1}
\end{figure*}

\begin{figure*}
    \includegraphics[width=11cm]{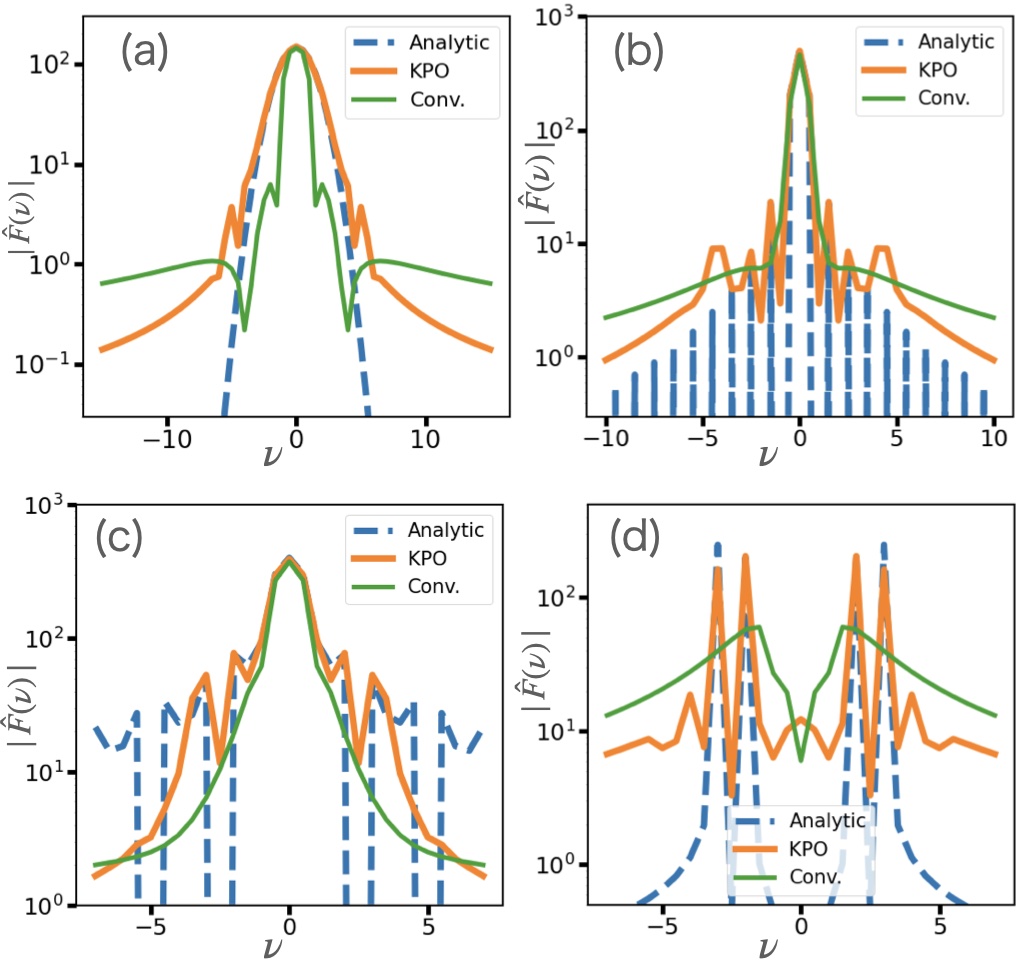}
    \caption{
     Plot of the absolute value of the Fourier transform of the function against the frequency.
     The blue dotted line denotes the function to be fitted.
     The orange (green) line denotes  the output by our (conventional) method after the optimization.
     The functions are
     (a) $e^{-36x^{2}}$, (b) $|x|$, (c) square wave and (d) $0.4\sin(4\pi x)+0.5\sin(6\pi x)$, respectively.}
        \label{fig2}
\end{figure*}

We show the results of the fitting in Fig.~\ref{fig1}. Our method approximates all functions better than the conventional method. In order to compare the expressibility more clearly, we define a Fourier transform as
\begin{align}
    \hat{F}(\nu)= \frac{1}{\sqrt{2\pi}}\int_{-1}^{1} dx F(x)e^{-2\pi i\nu x},
\end{align}
for any function $F(x)$, and we plot the absolute value of $\hat{f}(\nu)$ in Fig.~\ref{fig2}. As can be easily seen in (b) and (d), the results by our method contains more Fourier components than that by the conventional method.

Also, we plot the value of the cost function after the optimization by our method, and compare this with the conventional method in Table~\ref{tab:c_func1}.

\begin{table}[htbp]
    \centering
    \caption{Finally obtained values of the cost function.}\label{tab:c_func1}
    \begin{tabular}{c|cc}
    \hline
       $\tilde{f}(x)$ & KPO & Conventional \\
       \hline
        $e^{-36x^{2}}$ & $1.016\times 10^{-4}$ & $2.220\times 10^{-2}$ \\
        $|x|$ & $3.388\times10^{-4}$ & $7.923\times 10^{-3}$ \\
        Square wave & $1.344\times 10^{-2}$ & $4.661\times 10^{-2}$ \\
        $0.4\sin(4\pi x) + 0.5\sin(6\pi x)$ & $1.693\times 10^{-2}$ & $1.562\times 10^{-1}$\\
        \hline
    \end{tabular}
\end{table}
\begin{table}[htbp]
    \centering
    \caption{Numbers of iterations}\label{tab:num_of_iterations}
    \begin{tabular}{c|cc}
    \hline
       $\tilde{f}(x)$ & KPO & Conventional \\
       \hline
        $e^{-36x^{2}}$ & $5982$ & $6062$ \\
        $|x|$ & $5981$ & $6129$ \\
        Square wave & $6015$ & $6083$ \\
        $0.4\sin(4\pi x) + 0.5\sin(6\pi x)$ & $6041$ & $6115$\\
        \hline
    \end{tabular}
\end{table}

Next, let us discuss the case 
of the KPO network for $d_x=d_y=1$. 
Here, we use $\chi_{1} = \chi_{2} = 1$, $J_{12} = -0.1$, $K=2$, and $t_{d} = \tau = 1$. Also, by choosing $D=6$, we set the total number of parameters as 36, which is equal to that of the single KPO.

We could choose $\hat{M}=
(\hat{a}_{1} + \hat{a}^{\dagger}_{1})\otimes (\hat{a}_{2} +\hat{a}^{\dagger}_{2})
$ for our numerical simulations.
However, it is not straightforward to measure such a non-local observable with the KPO. So, instead,   we consider two observables $\hat{M}_{1} = \hat{a}_{1} + \hat{a}^{\dagger}_{1}$ and $\hat{M}_{2} = \hat{a}_{2} + \hat{a}^{\dagger}_{2}$. Also, we represent the function as $f(x;\boldsymbol{\theta})=\braket{\hat{M}_{1}}\braket{\hat{M}_{2}}$. We perform the fitting of the Gaussian and square wave, which we used in the case of the single KPO. Finally, we set the Hilbert-space cutoff dimension of each KPO as $10$. 

\begin{figure*}
    \centering
    \includegraphics[width = 11cm]{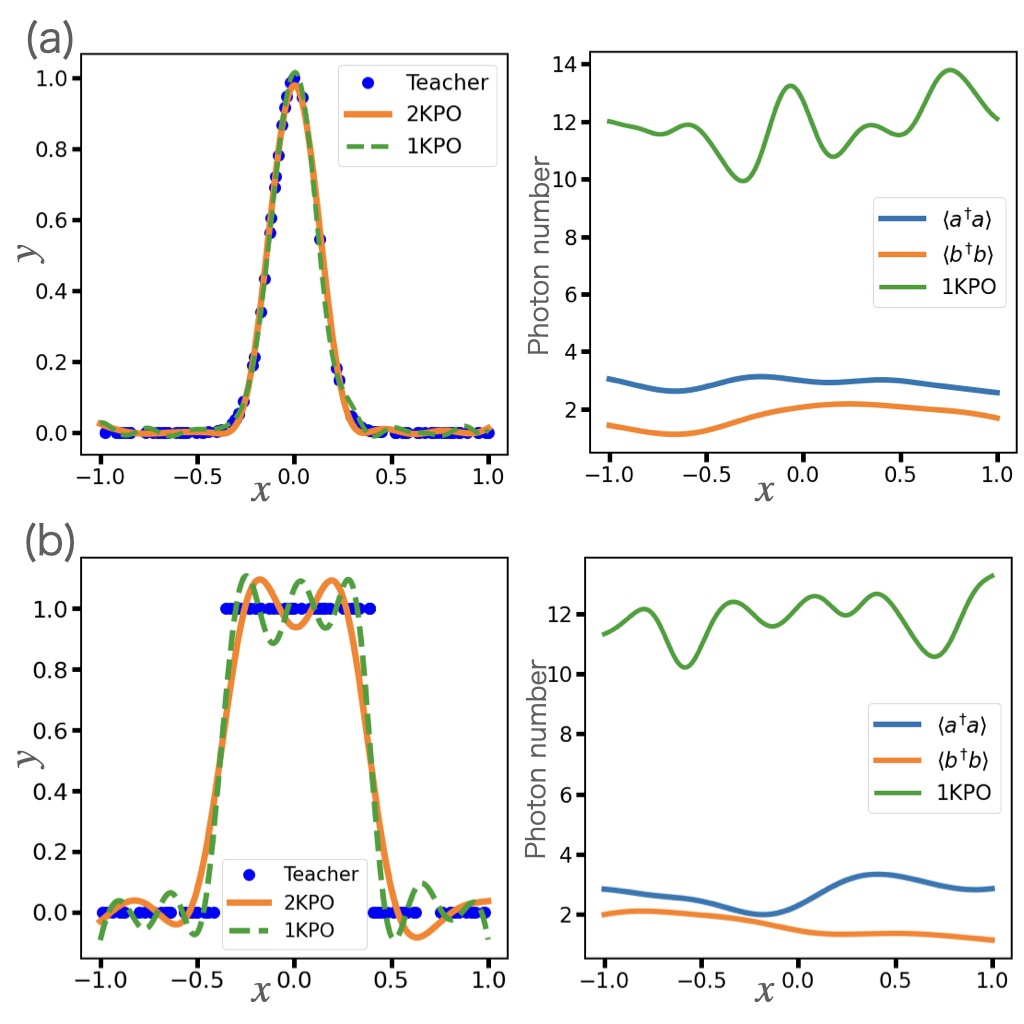}
    \caption{Demonstration results of our quantum machine learning for $e^{-36x^{2}}$ (a) and square wave (b) for the 1KPO and 2KPO cases. Left: the teacher data and the training results. Right: the average photon number. (Since we evaluated $\braket{a^{\dagger}a}$, the result depends on the variable $x$) }
    \label{fig:dkpo}
\end{figure*}

We plot the results in Fig.~\ref{fig:dkpo}, and compare the performance of our method to use the KPO network with that to use the single KPO. The cost functions after the optimization for 1KPO (2KPO) are $1.016\times 10^{-4}$ ($9.711\times 10^{-5}$) for the Gaussian ($e^{-36x^{2}}$) and $1.344\times 10^{-2}$ ($2.119\times 10^{-2}$) for the square wave. 
The performance of our method using the KPO network is similar to that using the single KPO.
However, we need to access higher excited states for the case of the single KPO than that of the KPO network, and therefore we could avoid the experimental difficulties by using the KPO network.

Let us explain the runtime of our scheme. During our simulations, we employed a maximum iteration count of 7200, the default setting provided by Scipy.optimize.minimize~\cite{scipy}, when dealing with $36$ variables. The optimization process terminates when the cost function either meets the predefined tolerance level (default value: $10^{-4}$) or when the maximum allowed number of iterations is reached. In either case, both the parameter set to minimize the cost function and the iteration number are outputted in Table~\ref{tab:num_of_iterations}.

In the KPO cases, the number of iterations is equal to or less than that in the conventional cases. The most time-consuming part of the practical runtime of the superconducting circuit is the execution time of two-qubit gates. 
Importantly, coupling strength between KPOs, as demonstrated in previous work~\cite{yamaji2022spectroscopic}, is approximately $10 \mathrm{MHz}$, which is similar to that of superconducting transmon qubits~\cite{stehlik2021tunable}.  Consequently, these findings highlight that the runtime of our method using KPOs is comparable with that of the conventional approach using transmon qubits.

\begin{figure*}
    \includegraphics[width=11cm]{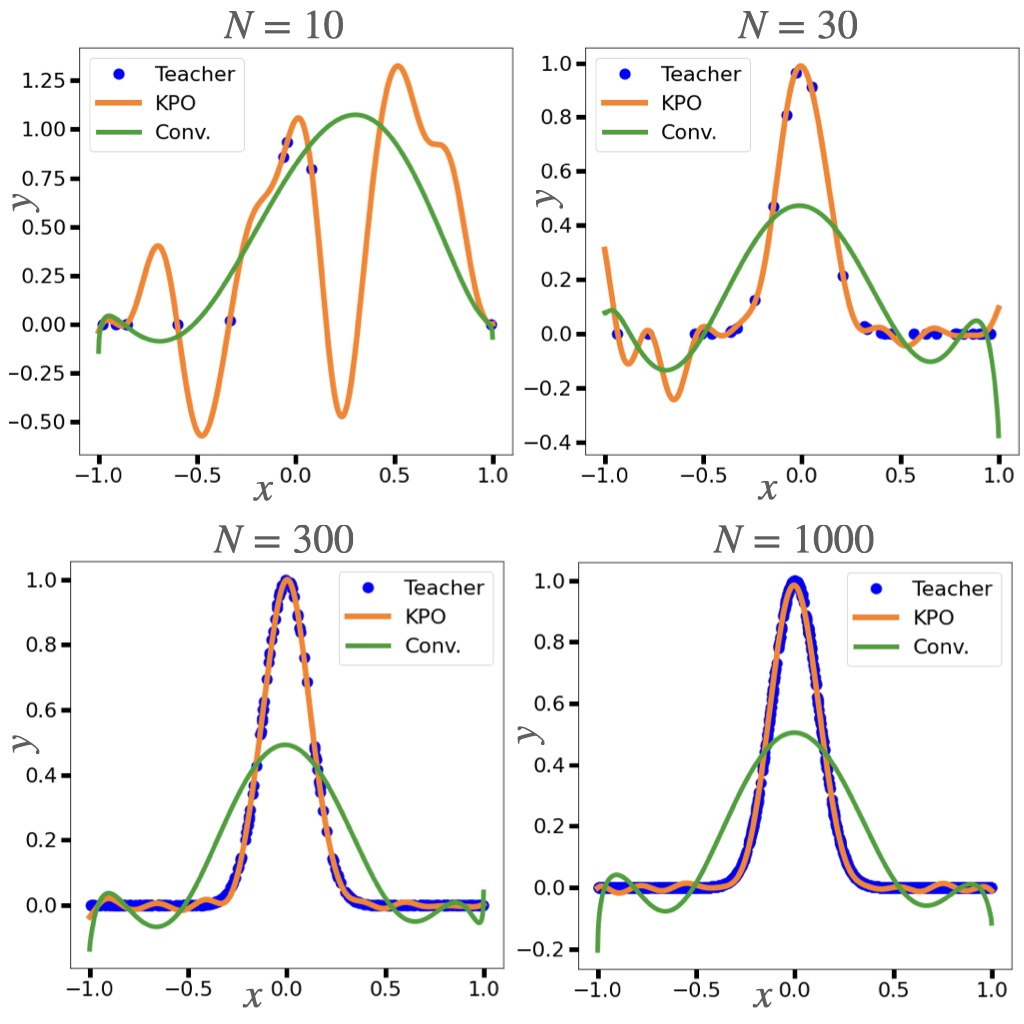}
    \caption{
        Results of our quantum machine learning for Gaussian described as $e^{-36x^{2}}$ with $N=10, 30, 300, 1000$ training data. When $N$ is small, overfitting occurs.}\label{sample_dep}
\end{figure*}

We show how our fitting results depend on the number of training data $N$ in Fig.~\ref{sample_dep} and the variation of the number of iterations in Table~\ref{tab:num_of_iterations2}.
For small $N$, our method seems to be susceptible to overfitting due to its inherent high expressiveness. 
Fortunately, to reduce the impact of overfitting, we can regulate this expressiveness by adjusting the photon number of the initial coherent state, as we will show in Sec.~\ref{subsec:alpha-expp}.

\begin{figure*}
    \includegraphics[width=11cm]{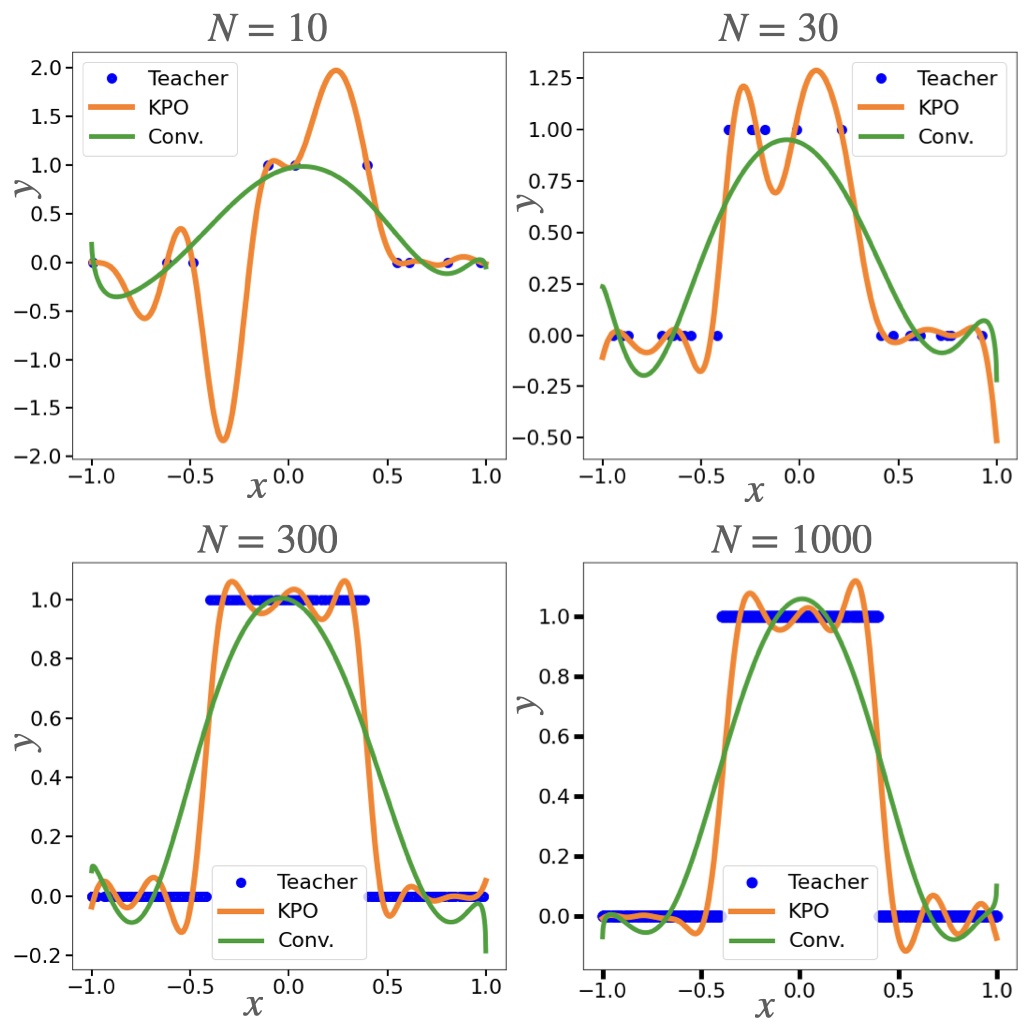}
    \caption{Results of our quantum machine learning for a square wave function with $N=10, 30, 300, 1000$ training data. When $N$ is small, overfitting occurs.}\label{sample_dep_sqwv}
\end{figure*}
\begin{widetext}
\begin{center}
\begin{table}[htbp]
    \caption{Variation of the number of iterations with the number of training data}\label{tab:num_of_iterations2}
    \begin{tabular}{c|cc|cc}
    \hline
       $N$ & KPO($e^{-36x^{2}}$) & Conventional($e^{-36x^{2}}$)&KPO(Square wave) &Conventional(Square wave) \\
       \hline
       $10$ & $4741$& $6050$ & $1821$  & $6084$\\
       $30$ & $6009$ & $6051$ & $5977$  & $6106$\\
        $100$ & $5982$& $6062$ & $6015$ & $6083$ \\
        $300$ & $6000$ & $6042$ & $6009$ & $6080$ \\
        $1000$ & $6008$ & $6062$& $6080$ & $6058$ \\
        \hline
    \end{tabular}
\end{table}
\end{center}
\end{widetext}
%%%%%
\subsection{$\alpha$ and expressive power}
\label{subsec:alpha-expp}
%%%%%

From Eq.~\eqref{eq:fourier_peak}, we find a tendency that, as we increase (decrease) $\alpha$, more (less) high frequency terms are added.  Therefore, it is expected that we can control the expressibility by tuning the size of the coherent state prepared as the initial state.

\begin{figure*}
    \centering
    \includegraphics[width = 11cm]{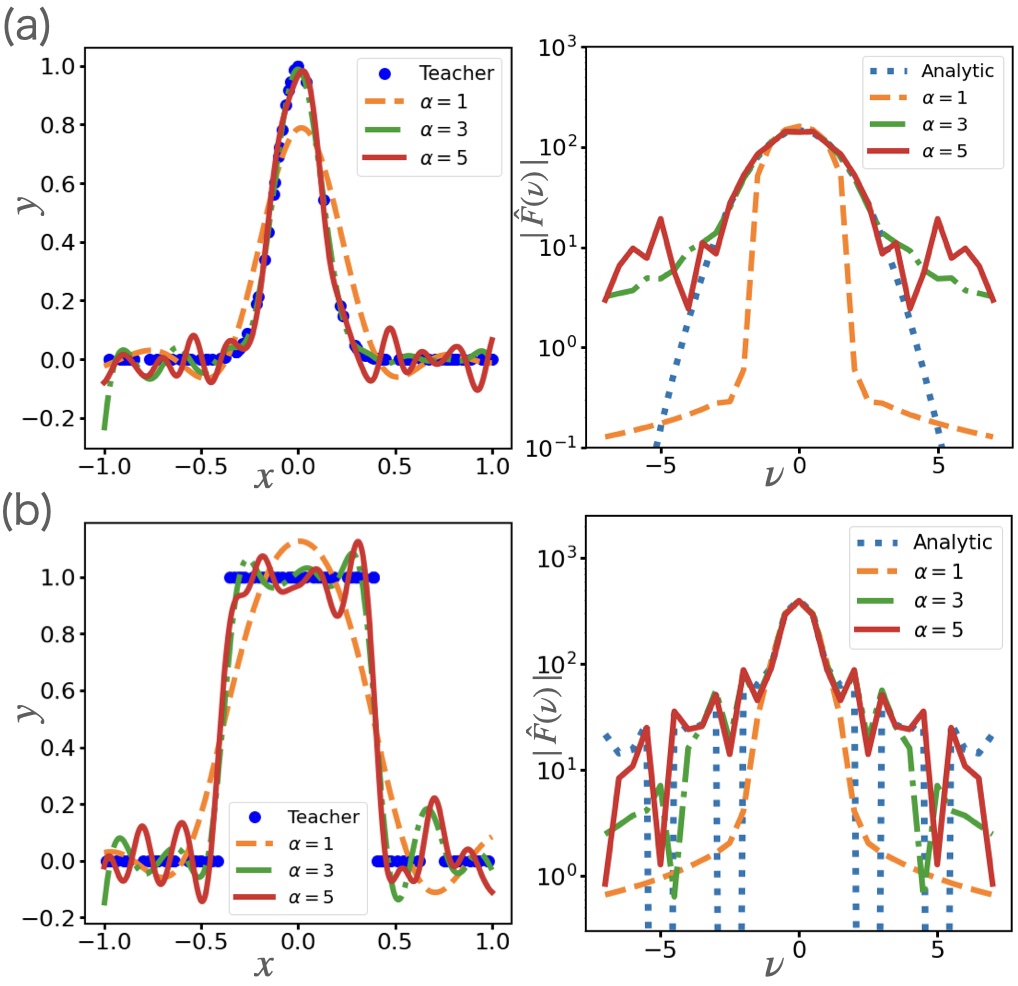}
    \caption{Demonstration results of our quantum machine learning for $e^{-36x^{2}}$ (a) and the square wave (b) for the different $\alpha=1, 3, 5$ cases. Left: the teacher data and the training results. Right: the Fourier spectrum of the training results.}
    \label{fig:alpha_dep}
\end{figure*}

We confirmed this point by numerical simulations. In Fig.~\ref{fig:alpha_dep}, we perform numerical simulations for $\alpha = 1, 3,$ and $5$ with the use of supervised data generated from two functions, a Gaussian 
and a square wave. 
Only for $\alpha = 5$, we change the cutoff dimension of the Hilbert space from $25$ to $100$ because the average photon number is $25$. 

In machine learning, there is a trade-off between increasing expressive power and overfitting. This means that, as we increase the expressibility, the problem of the overfitting becomes more severe. In our method, we could tune the parameter $\alpha$ to choose the best point for the fitting.

\begin{figure}
    \includegraphics[width = 8.7cm]{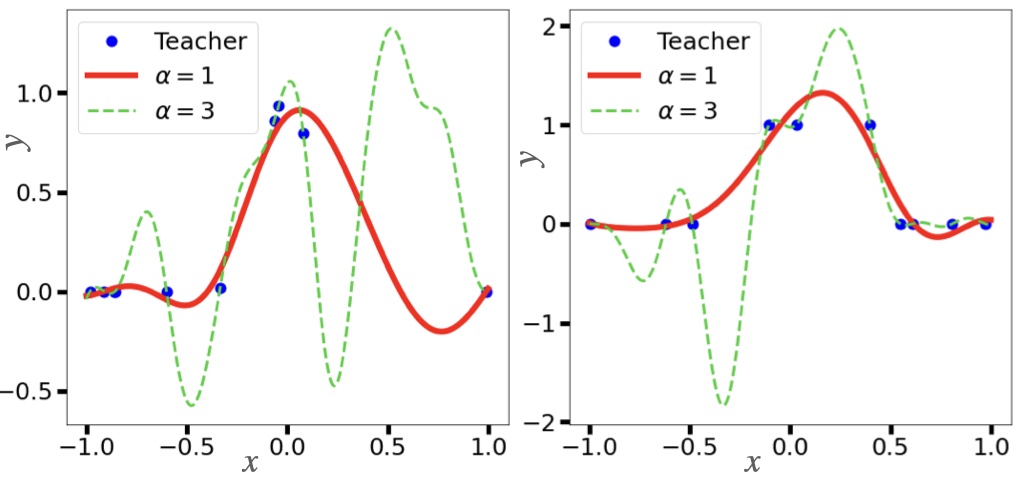}
    \caption{
    Results of our quantum machine learning for Gaussian (left side) and the square wave function (right) for $N=10$ where we change the photon number of the initial coherent state. We successfully mitigate the impact of the overfitting by reducing the photon number of the coherent state.
    }\label{av_ov}
\end{figure}

To illustrate our concept, we performed numerical simulations in which we varied the number of photons in the initial coherent state. 
As we mentioned before, in Fig.~\ref{sample_dep}, overfitting occurs for a smaller number of the training data $N$. We apply our method to tune the expressibility to this case. 
In Fig.~\ref{av_ov}, 
we present the results, highlighting that reducing the photon number of the initial coherent state effectively mitigates the impact of overfitting.

%%%%%%%%%%%%%%
\section{Conclusions and Discussion}
\label{sec:conclusion_and_discussion}
%%%%%%%%%%%%%%
In conclusion, we propose to use the KPO for the quantum supervised machine learning with variational quantum circuits. We numerically show that, although we use a single KPO, the expressibility of our method is higher than the conventional method with six qubits.
In our method, we can tune an amplitude of the initial coherent state, and we numerically show that the expressibility increases as we increase the amplitude.

In this paper, we provide proof of concept using a regression problem as an example. Our method also offers advantages due to its expressive nature for other machine learning problems, including classification, generation, reinforcement learning, and sequential learning. Furthermore, we acknowledge that the quantum kernel method~\cite{havlivcek2019supervised} could be another promising application of our approach, as our data encoding methodology into quantum states introduces new types of quantum kernels. Exploring these applications is a promising direction for future research.

In the NISQ era, it is crucial to implement the algorithm with a fewer resource, and our results to use the KPO will contribute to reduce resource. KPO network may be used as a variant of continuous variable neural network~\cite{Killoran_2019PRR}. 
There are many potential applications to use the continuous degrees of freedom of the KPO. We hope that our research will help to expand the range of applications of the KPO.

\begin{acknowledgments}
We would like to thank Takashi Imoto, Suguru Endo, Takaaki Aoki for fruitful discussions. 
This work was supported by the Leading Initiative for Excellent Young Researchers, MEXT, Japan, and JST Presto (Grant No.~JPMJPR1919), Japan. This work was also supported by Grant-in-Aid for JSPS Research Fellow 22J01501. This paper is partly based on the results obtained from a project, JPNJ16007, commissioned by the New Energy and Industrial Technology Development Organization (NEDO), Japan. 
This
work was supported by JST Moonshot R\&D (Grant Number JPMJMS226C).
We also would like to thank the developers of QuTiP~\cite{Qutip} and Qulacs~\cite{Qulacs}. They were used for our numerical simulations.
\end{acknowledgments}

%%%%%%%%%%
\subsection*{Declarations}
\paragraph*{Competing interests}
Not applicable.
\paragraph*{Authors' contributions}
Y.Mori performed the calculations and wrote the draft of the manuscript. Also, Y.Mori prepared the figures. K.N. introduced the motivation of this study and wrote the abstract and introduction in terms of studies of the VQA. Y.Matsuzaki and S.K. supervised the project.  All authors reviewed and revised the manuscript.
\paragraph*{Availability of data and materials}
The datasets used and/or analyzed during the current study are available from Y.Mori on reasonable request.
%%%%%%%%%%

\appendix

%%%%%%%%%%
\section{A derivation of a Hamiltonian of a KPO}
\label{append:KPO_principle}
%%%%%%%%%%

We explain a derivation of the effective Hamiltonian of the KPO composed of an array of superconducting quantum interference devices (SQUIDs)~\cite{wang2019quantum, Gao_2021PRXQ, Masuda_2022PRAppl}.
Let us denote the overall phase across the junction array by $\hat{\varphi}$. Also, we denote $\hat{n}_{C}$ by a difference in the number of charges on the two plates of the capacitor in units $2e$, which corresponds to the charge of a Cooper pair.
\begin{figure}
    \centering
    \includegraphics[width = 6cm]{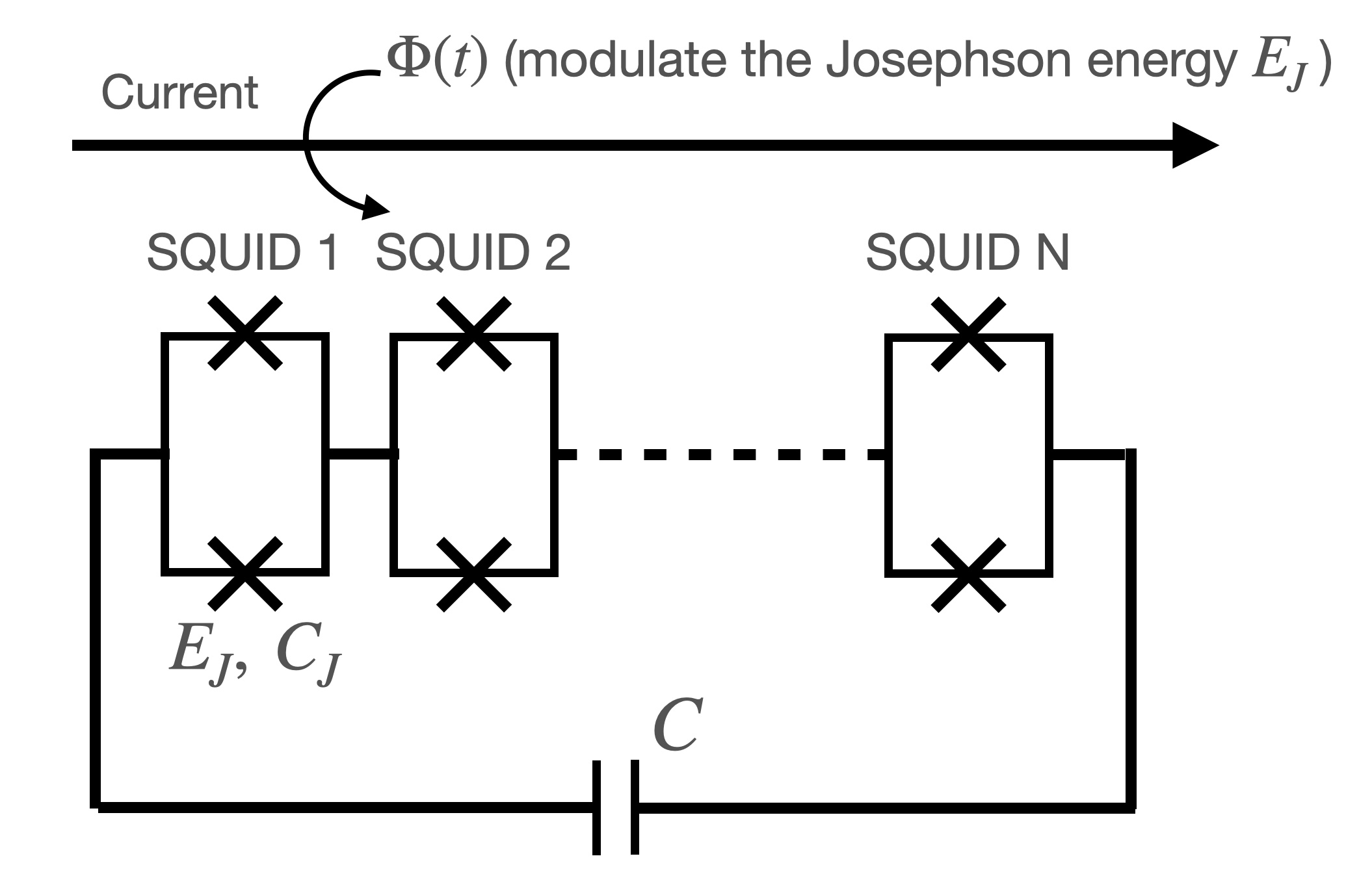}
    \caption{Circuit model of a KPO with $N$ SQUIDs and a shunt capacitor $C$. }
    \label{squidfig}
\end{figure}
These will obey the following canonical commutation relation.
\begin{align}
    [\hat{\varphi},\hat{n}_{C}]=i.\label{can_qua}
\end{align}
The Hamiltonian of the system is given as
\begin{align}
    \hat{H} =4 E_{C}\hat{n}_{C}^{2} -N_{S}E_{J}[\Phi(t)]\cos{\frac{\hat{\varphi}}{N_{S}}},\label{Joseph_Ham}
\end{align}
where $N_{S}$, $E_{J}$, and $E_{C}$ are the number of SQUIDs, and the Josephson energy of a SQUID, and the charging energy of the resonator.

By using a time-dependent magnetic flux $\Phi(t)$ penetrating the SQUID loops, we can modulate the Josephson energy as $E_{J} (t) = E_{J} + \delta E_{J}  \cos{\omega_{p}t}$. For simplicity, we set the phase of the pump field to be zero, $\theta = 0$.

By performing a Taylor expansion and taking into account up to the fourth order of $\varphi/N_{S}$, we approximate Hamiltonian as
\begin{align}
    \hat{H}&=4E_{C}\hat{n}_{C}^{2}+\frac{E_{J}}{2N_{S}}\hat{\varphi}^{2}-\frac{E_{J}}{24N_{S}^{3}}\varphi^{4}\nonumber\\
    &\quad -N_{S}\delta E_{J}\cos{\omega_{p}t}\left(1-\frac{1}{2}\frac{\hat{\varphi}^{2}}{N_{S}^{2}}+\frac{1}{24}\frac{\hat{\varphi}^{4}}{N_{S}^{4}}\right).\label{TalHam}
\end{align}
Here, we assume $|\langle \varphi/N_{S}\rangle| \ll 1$.

We can define the creation and annihilation operator as follows.
\begin{align}
    \hat{a} &= \frac{1}{2}\left(\frac{E_{J}}{2N_{S}E_{C}}\right)^{\frac{1}{4}}\hat{\varphi} + i\left(\frac{2N_{S}E_{C}}{E_{J}}\right)^{\frac{1}{4}}\hat{n}_{C},\\
    \hat{a}^{\dagger} &= \frac{1}{2}\left(\frac{E_{J}}{2N_{S}E_{C}}\right)^{\frac{1}{4}}\hat{\varphi} - i\left(\frac{2N_{S}E_{C}}{E_{J}}\right)^{\frac{1}{4}}\hat{n}_{C}.
\end{align}
Then, the Hamiltonian~\eqref{TalHam} becomes
\begin{align}
    \hat{H}&=\omega\left(\hat{a}^{\dagger} \hat{a} +\frac{1}{2}\right)-\frac{K}{12}(\hat{a}+\hat{a}^{\dagger})^{4}\nonumber\\
    &\quad +\left[-N_{S}\delta E_{J} +p(a+a^{\dagger})^{2} -\frac{Kp}{3\omega}(a+a^{\dagger})^{4}\right]\nonumber\\
    &\qquad \times \cos{\omega_{p}t},\label{hamilt_ve}
\end{align}
where $\omega=\sqrt{8E_{C}E_{J}/N_{S}}$, $K=E_{C}/N_{S}^{2}$, and $p=2\omega\delta E_{J}/8E_{J}$. 

We assume $Kp\ll \omega$, and we drop the last term in Eq.~\eqref{hamilt_ve}. Then, we obtain
\begin{align}
    \hat{H} = \omega a^{\dagger}a -\frac{K}{12}(a+a^{\dagger})^{4} + p(a+a^{\dagger})^{2}\cos{\omega_{p}t}
\end{align}
Moving to the rotating frame at the frequency of $\omega_{p}/2$ and using the rotating wave approximation, we obtain
\begin{align}
    \hat{H}=\Delta a^{\dagger} a -\frac{K}{12}a^{\dagger 2}a^{2}+\frac{p}{2}(a^{2}+a^{\dagger 2}),
\end{align}
where $\Delta = \omega-K-\omega_{p}/2.$

%%%%%%%%%%
\section{A conventional scheme for our simulation for $d_{x}=d_{y}=1$}
\label{app:A}
%%%%%%%%%%
Let us review a conventional scheme for quantum circuit learning with qubits~\cite{Mitarai_2018_PRA}.
The unitary gate to encode input data $U(\boldsymbol{x})$ is chosen as
\begin{align}
 \hat{U}(x)=\prod_{j=1}^{M}R^{Z}_{j}(\arccos(x^{2}))R^{Y}_{j}(\arcsin(x)),
\end{align}
where $R^{Z}_{j}(\phi)$ is the rotation of the $j$-th qubit around the $z$ axis with an angle of $\phi$ and $R^{Y}_{j}(\phi)$ is the rotation of the $j$-th qubit around the $y$ axis with an angle of $\phi$, respectively and $M$ is the number of the qubits.

The parameterized unitary $\hat{V}(\boldsymbol{\theta})$ is composed of two parts. The first part is a single qubit rotation on the $j$-th qubit $\hat{U}(\theta^{i}_{j})$, which can be generally decomposed into the following form
\begin{align}
\hat{U}(\theta^{i}_{j}) = R^{X}_{j}(\theta^{i}_{j1})R^{Z}_{j}(\theta^{i}_{j2})R^{X}_{j}(\theta^{i}_{j3}).
\end{align}
This means that the single-qubit rotation gate contains three free parameters. The other part is a unitary operation induced by the following Hamiltonian 
\begin{align}
\hat{H} = \sum_{j=1}^{M} a_{j}X_{j} +\sum_{j=1}^{M}\sum_{k=1}^{j-1}J_{jk}Z_{j}Z_{k},
\end{align}
where the coefficients $a_{j}$ and $J_{jk}$ are taken randomly from a uniform distribution on $[-1, 1]$.
This means that the unitary operator becomes
\begin{align}
    \hat{V}(\boldsymbol{\theta}) = \prod_{i=1}^{D}e^{-i\tau \hat{H}}\prod_{j = 1}^{K}\hat{U}(\theta^{i}_{j}).\label{op:v_ordin}
\end{align}
We use $\tau = 10$ in our simulation, 
which is the same as that used in~\cite{Mitarai_2018_PRA}.

%%%%%%%%%%
\section{A method for supervised machine learning with $d_x>1$ by using a single KPO}
\label{append:mulplefun1kpo}
%%%%%%%%%%
We could use the single KPO with $d_x>1$ and $d_y>1$.
For $d_x>1$, we need to encode the variable  into the initial state.  
If we encode a different data, different outputs should be generated. 
So, in this case, $\ket{\psi(x_1,.... .x_{d_{x}})}$ should be different from
$\ket{\psi(x'_1,.... .x'_{d_{x}})}$ unless $x_j=x'_j$ for $j=1,2,\cdots, d_x$ is satisfied.
For $d_y>1$, we have to measure independent observables $\hat{M}_{1}$, $\hat{M}_{2}$, ..., $\hat{M}_{d_{y}}$ at the step 3.

Let us explain such an example.
Firstly, we prepare the coherent state $\ket{\alpha}$ where we set $\alpha = r = \sqrt{x_{1}^{2}+x_{2}^{2}}$. Second, we perform a unitary operation $e^{-i\tilde{\chi}\hat{n}^{2}-i\varphi\hat{n}}$, where $\varphi$ is defined by
\begin{align}
    \varphi = 
    \begin{cases}
    \arccos{\frac{x_{1}}{r}},\quad (x_2 > 0) \\
    -\arccos{\frac{x_{1}}{r}},\quad (x_2 \leq 0) \\
    0,\quad (r = 0).
    \end{cases}
\end{align}
It is notable that $\varphi$ is defined on the interval $[-\pi, \pi).$
We obtain 
\begin{align}
    \ket{\psi(x_{1},x_{2})}=e^{-i\tilde{\chi}\hat{n}^{2}-i\varphi\hat{n}}\ket{r}.
\end{align}

The overlap $\braket{\psi(x'_{1},x'_{2})|\psi(x_{1},x_{2})}$ is calculated as follows
\begin{align}
&\braket{\psi(x'_{1},x'_{2})|\psi(x_{1},x_{2})}\nonumber\\
&=\bra{r'}e^{i\tilde{\chi}\hat{n}^{2}+i \varphi'\hat{n}}e^{-i\tilde{\chi}\hat{n}^{2}-i \varphi\hat{n}}\ket{r}\nonumber\\
&=\bra{r'}e^{i(\varphi'-\varphi)\hat{n}}\ket{r}\nonumber\\
&=\braket{r'|re^{i(\varphi'-\varphi)}}\nonumber\\
&=e^{-\frac{1}{2}(|r'|^{2}+|r|^{2}-2r'r e^{i(\varphi'-\varphi)})}.\label{eq:ovl_dx1}
\end{align}
If the overlap \eqref{eq:ovl_dx1} is $1$, the exponent of the overlap is zero, and we obtain
\begin{align}
    r'^{2}+r^{2}-2r'r e^{i(\varphi'-\varphi)} = 0.
\end{align}
We obtain the solution as follows.
\begin{align}
    \varphi'&= \varphi,\\
    r' &= r.
\end{align}
Therefore, the overlap is not unity unless $x_1=x_1'$ and $x_2=x_2'$ are satisfied.
Also, as the observable, we can adopt $\hat{M}_{1}=\hat{a}+\hat{a}^{\dagger}$
and $\hat{M}_{2}=\hat{a}^{\dagger}\hat{a}$.
So we can use this initial state for our method with $d_x=2$ and $d_y=2$.

\bibliography{main}

\end{document}